\documentclass[apj]{emulateapj}
\usepackage[colorlinks=true,linkcolor=blue,citecolor=blue]{hyperref}

\usepackage{latexsym,amssymb}
\usepackage{amsmath}
\usepackage{natbib}

\usepackage{color}
\usepackage{graphics}

\newcommand{\mysim}{\mathord{\sim}}
\newcommand{\mylesssim}{\mathord{\lesssim}}
\newcommand{\mygtrsim}{\mathord{\gtrsim}}
\newcommand{\myapprox}{\mathord{\approx}}

\def\be{\begin{eqnarray}}
\def\ee{\end{eqnarray}}
\def\no{\nonumber}

\newcommand{\frot}{f_{\textrm{rot}}}
\newcommand{\frotz}{f_{\textrm{rot},0}}

\begin{document}
\title{Neutrino Signal of Collapse-Induced Thermonuclear Supernovae: The Case for Prompt Black Hole Formation in SN1987A} 

\author{ Kfir Blum$^{1,2,3}$, Doron Kushnir$^{2,3}$}
\affiliation{$^1$Department of Particle Physics and Astrophysics, Weizmann Institute of Science, Rehovot, 760000 Israel}
\affiliation{$^2$School of Natural Sciences, Institute for
Advanced Study, Princeton, New Jersey, 08540 USA}
\affiliation{$^3$John N.\ Bahcall Fellow}
\email{$^a$kfir.blum@weizmann.ac.il}
\email{$^b$kushnir@ias.edu}
%\date{\today}

\begin{abstract}
Collapse-induced thermonuclear explosion (CITE) may explain core-collapse supernovae (CCSNe). We present a first, preliminary analysis of the neutrino signal predicted by CITE and compare it to the neutrino burst of SN1987A. For strong ($\mygtrsim10^{51}$ erg) CCSNe, as SN1987A, CITE predicts a proto-neutron star (PNS) accretion phase, accompanied by the corresponding neutrino luminosity, that can last up to a few seconds and that is cut-off abruptly by a black hole (BH) formation. The neutrino luminosity can later be revived by accretion disc emission after a dead time of few to a few ten seconds. In contrast, the competing neutrino mechanism for CCSNe predicts a short ($\lesssim$ sec) PNS accretion phase, followed, upon the explosion, by a slowly declining PNS cooling luminosity. We repeat statistical analyses used in the literature to interpret the neutrino mechanism, and apply them to CITE. The first 1-2 sec of the neutrino burst are equally compatible with CITE and with the neutrino mechanism. However, the data hints towards a luminosity drop at t=2-3 sec, that is in some tension with the neutrino mechanism while being naturally attributed to BH formation in CITE. The occurrence of neutrino signal events at 5 sec in SN1987A places a constraint on CITE, suggesting that the accretion disc formed by that time. We perform 2D numerical simulations, showing that CITE may be able to accommodate this disc formation time while reproducing the ejected $^{56}$Ni mass and ejecta kinetic energy within factors 2-3 of observations. We estimate the accretion disc neutrino luminosity and show that it can roughly match the data. This suggests that direct BH formation is compatible with the neutrino burst of SN1987A. With current neutrino detectors, the neutrino burst of the next strong Galactic CCSN may give us front-row seats to the formation of an event horizon in real time. Access to phenomena near the event horizon motivates the construction of a few Megaton neutrino detector that should observe extragalactic CCSNe on a yearly basis.
\end{abstract}

\maketitle

%\tableofcontents
%%%%%%%%%%%%%%%%%%%
\section{Introduction}
There is strong evidence that type-II supernovae (SNe) are explosions of massive stars, initiated by the gravitational collapse of the stars' iron core \citep[][]{Burbidge:1957vc,Hirata:1987,Smartt:2009zr}. It is widely thought that the explosion is obtained due to the deposition in the envelope of a small fraction ($\mysim1\%$) of the gravitational energy ($\mysim10^{53}\,\textrm{erg}$) released in neutrinos from the core, leading to the $\mysim10^{51}\,\textrm{erg}$ observed kinetic energy ($E_{\textrm{kin}}$) of the ejected material \citep[see][for reviews]{Bethe:1990,Janka:2012}. However, one-dimensional (1D) simulations indicate that the neutrinos do not deposit sufficient energy in the envelope to produce the typical $E_{\textrm{kin}}\mysim10^{51}\,\textrm{erg}$. While some two-dimensional (2D) studies indicate successful explosions \citep[][]{Bruenn:2013,Bruenn:2014,2015PASJ..tmp..261N,Suwa}, others indicate failures or weak explosions \citep[][]{Takiwaki:2014,Dolence:2015}, and these studies are affected by the assumption of rotational symmetry and by an inverse turbulent energy cascade that, unlike many physical systems, appears to amplify energy on large scales. Therefore, three-dimensional (3D) studies are necessary to satisfactorily demonstrate the neutrino mechanism, but so far 3D studies have resulted in either failures or weak explosions \citep[][]{Takiwaki:2014,Lentz:2015,Melson:2015a,Melson:2015b}. 

\citet[][]{Burbidge:1957vc} suggested a different mechanism for the explosion that does not involve the emitted neutrinos. They suggested that the adiabatic heating of the outer stellar shells as they collapse triggers a thermonuclear explosion \citep[see also][]{Hoyle:1960,Fowler:1964}. This collapse-induced thermonuclear explosion (CITE) has the advantage of naturally producing $E_{\textrm{kin}}\mysim10^{51}\,\textrm{erg}$ from the thermonuclear burning of $\mysim M_{\odot}$ of light elements, with a gain of $\mysim \textrm{MeV}$ per nucleon. A few 1D studies suggested that this mechanism does not lead to an explosion because the detonation wave is ignited in a supersonic in-falling flow \citep[][]{Colgate:1966,Woosley:1982,Bodenheimer:1983}, and the idea was subsequently abandoned. While the results of these studies are discouraging, they only demonstrate that some specific initial stellar profiles do not lead to CITE, and they do not prove that CITE is impossible for all profiles.

Recently, \citet[][]{Kushnir:2014oca} have shown that CITE is possible in some (tuned) 1D initial profiles, that include shells of mixed helium and oxygen, but resulting in weak explosions, $E_{\textrm{kin}}\mylesssim10^{50}\,\textrm{erg}$, and negligible amounts of ejected $^{56}$Ni. Subsequently, \citet{Kushnir:2015mca} used 2D simulations of rotating massive stars to explore the conditions required for CITE to operate successfully. It was found that for stellar cores that include slowly (a few percent of breakup) rotating shells of mixed He-O with densities of $\textrm{few}\times10^{3}\,\textrm{g}\,\textrm{cm}^{-3}$, a thermonuclear detonation that unbinds the stars' outer layers is obtained. With a series of simulations that cover a wide range of progenitor masses and profiles, it was shown that CITE is insensitive to the assumed profiles and thus a robust process that leads to supernova explosions for rotating massive stars. The resulting explosions have $E_{\textrm{kin}}$ in the range of $10^{49}-10^{52}\,\textrm{erg}$ and ejected $^{56}$Ni masses ($M_{\textrm{Ni}}$) of up to $\mysim1\,M_{\odot}$, both of which cover the observed ranges of core-collapse supernovae (CCSNe, including types II and Ibc). 
 
It is difficult to test observationally if the initial conditions required for CITE exist in nature. Nevertheless, CITE makes a few predictions that are different from the predictions of the neutrino mechanism, and that can be compared to observations. For example, CITE predicts that stronger explosions (i.e., larger $E_{\textrm{kin}}$ and higher $M_{\textrm{Ni}}$) are obtained from progenitors with higher pre-collapse masses. \citet{Kushnir:2015vka} showed that the observed correlation between $M_{\textrm{Ni}}$ and the luminosities of the progenitors for type II SNe is in agreement with the prediction of CITE and in possible contradiction with the neutrino mechanism. Another prediction of CITE is that neutron stars (NSs) are produced in weak ($E_{\textrm{kin}}\mylesssim10^{51}\,\textrm{erg}$) explosions, while strong ($E_{\textrm{kin}}\mygtrsim10^{51}\,\textrm{erg}$) explosions leave a black hole (BH) remnant. This prediction suggests that a BH was formed in SN1987A \citep[$E_{\textrm{kin}}\approx1.5\cdot10^{51}\,\textrm{erg}$;][]{Utrobin:2011} during the first few seconds after core collapse (direct BH formation, to be distinguished from BH formation from fallback, which lasts hours to days). In contrast, simulations based on artificially triggered explosions within the neutrino mechanism predict that the compact object in SN1987A is a NS \citep[see, e.g.,][]{Perego:2015oqa}. At the time of writing, a NS has not yet been detected in the cite of SN1987A~\citep{Graves:2005xy,Larsson:2011hb}, but see \citet{Zanardo:2014uoa} for a possible recent hint.

In this paper we continue to explore the observational consequences of CITE. We focus on the neutrino signal characterizing core-collapse, and derive constraints from the neutrino burst that accompanied SN1987A~\citep{Bionta:1987qt,Hirata:1987hu}. Specifically, we ask, and begin to answer, the following two questions.
\begin{enumerate}
\item As mentioned above, CITE predicts that a BH was formed directly during the event of SN1987A. The reason for this expectation is that a strong explosion ($E_{\textrm{kin}}\mygtrsim10^{51}\,\textrm{erg}$) requires a high mass for the He-O shell ($\mygtrsim1\,M_{\odot}$), which in turn requires a massive core ($\mygtrsim4\,M_{\odot}$) below the shell. By the time CITE operates (on the order of the free-fall time of the He-O layer $\mysim30\,\textrm{sec}$) the mass below the He-O shell accrets onto the central proto-neutron star (PNS), topping the critical mass and turning it into a BH \citep[within 1-3~sec;][]{OConnor:2011}.

Direct BH formation in SN1987A has been previously considered unlikely, as it was argued to abruptly terminate the PNS neutrino emission\footnote{Another frequent argument in the literature~\citep[see, e.g.,][]{Mirizzi:2015eza} is that BH formation would not be compatible with the neutrino mechanism for the explosion. This argument is of course irrelevant to our analysis here.}. This would be incompatible with the detection of neutrinos at later times \citep[][]{Burrows:1988,Loredo:2001rx}: as we review below, neutrino signal events were detected 5-10~sec after core-collapse. Our first question is: does this argument rule out CITE?

We show here that the answer is negative. CITE, and more generally direct BH formation, can be reconciled with the SN1987A neutrino signal. Even though BH formation should indeed temporarily quench the neutrino burst, the subsequent formation of an accretion disc around the BH can produce a neutrino luminosity consistent with observations. 

The fact that accretion disc during stellar collapse can produce the required late-time neutrino luminosity should come as no surprise. Similar scenarios have been investigated in the literature in the context of the collapsar model for gamma-ray bursts~\citep[GRBs;][]{MacFadyen:1998vz}, and the resulting discs have been shown to exhibit copious neutrino emission\footnote{See also~\cite{Liu:2015prx} for a recent analysis.}~\citep{Popham:1998ab,MacFadyen:1998vz}. It is interesting to note that \citet[][]{Loredo:2001rx} in their analysis of SN1987A found that direct BH formation is favored by the neutrino data, but they set a prior against this possibility. We repeat here a similar likelihood analysis of SN1987A and show that CITE can indeed give a somewhat better fit to the  data. 

\item A key to CITE is the formation of a rotationally-induced accretion shock (RIAS) during the collapse of the stellar envelope below the He-O layer~\citep{Kushnir:2015mca}. The RIAS provides the match for thermonuclear explosion. Importantly for us here, the RIAS formation time is precisely the formation time of the accretion disc that is needed to restart the neutrino luminosity after BH formation.
As mentioned above, SN1987A data implies that the accretion disc neutrino luminosity should be operative by $t\sim5$~sec. Our second question is: can CITE operate successfully with RIAS formation time as early as a few seconds? 

\citet{Kushnir:2015mca} made preliminary studies of the dependence of CITE on the pre-collapse 
stellar profile, but for profiles which resulted in strong explosions ($E_{\textrm{kin}}>10^{51}\,\textrm{erg}$) the RIAS formation times considered there were  significantly larger than $5\,\textrm{sec}$. Here we extend the analysis of~\cite{Kushnir:2015mca} by further numerical simulations. Guided by the neutrino data of SN1987A we tailor the initial profile to initiate neutrino emission from a disc at $t_{\textrm{disc}}\approx5$~sec. With this $t_{\textrm{disc}}$ constraint we are able to find a profile in which CITE operates and yields values of $M_{\textrm{Ni}}\sim0.035~{\rm M_{\odot}}$ and $E_{\textrm{kin}}\sim0.6\cdot10^{51}$~erg, in the ballpark of, though a factor 2-3 below observations for SN1987A~\citep{Hamuy:2002qx,Utrobin:2011}. 

We further estimate the neutrino emission from the accretion disc at the base of the RIAS, finding a $\bar\nu_e$ luminosity of about $L_{\bar\nu_e}\sim0.5\cdot10^{51}$~erg/sec and mean neutrino energy $\sim10$~MeV. These results are on the low side, but not inconsistent with the range allowed by the data. While more simulations are needed for conclusive results, our preliminary findings indicate that CITE can operate in rough agreement with the neutrino data of SN1987A.
\end{enumerate}

In Section~\ref{sec:data} we review the neutrino light curve from SN1987A, recalling the  signal events at $5-10$~sec that place an important constraint on CITE. We further note a hint for a drop in $\bar\nu_e$ luminosity around $t\sim2$~sec, and examine it in Section~\ref{ssec:lumin1}. While the luminosity drop is not very statistically significant, we find it interesting to repeat a likelihood analysis as of~\citet{Loredo:2001rx} and \citet{Pagliaroli:2008ur} for the neutrino mechanism, but focusing on interpretation within CITE. In Section~\ref{ssec:lumin2} we show that the neutrino mechanism is in some tension with the luminosity drop, of order two standard deviations. In Section~\ref{ssec:lumin3} we give a toy model parameterization of the neutrino burst expected in CITE, with the same number of free parameters as used in~\citet{Loredo:2001rx} and in~\citet{Pagliaroli:2008ur} to describe the neutrino mechanism. The CITE model is fitted to data with reasonable parameters and naturally addresses the luminosity drop at $t\sim2$~sec. 

In Section~\ref{sec:numerical} we use 2D numerical simulations to demonstrate that CITE can operate successfully with early RIAS formation time $t_{\textrm{disc}}\sim5$~sec, yielding $E_{\textrm{kin}}$ and $M_{\textrm{Ni}}$ in rough agreement with SN1987A. We show that the BH accretion disc at the base of the RIAS can revive the neutrino emission with luminosity in the ballpark seen in the data. We conclude in Section~\ref{sec:conclusion}. In Appendix~\ref{app:EC} we recap some details of the phenomenological modelling of the neutrino mechanism and add some more statistical analyses.

%%%%%%%%%%%%%%%%%%%%%%%%
\section{SN1987A neutrino data}
\label{sec:data}
\begin{figure}[htbp]
\includegraphics[width=1.1\linewidth]{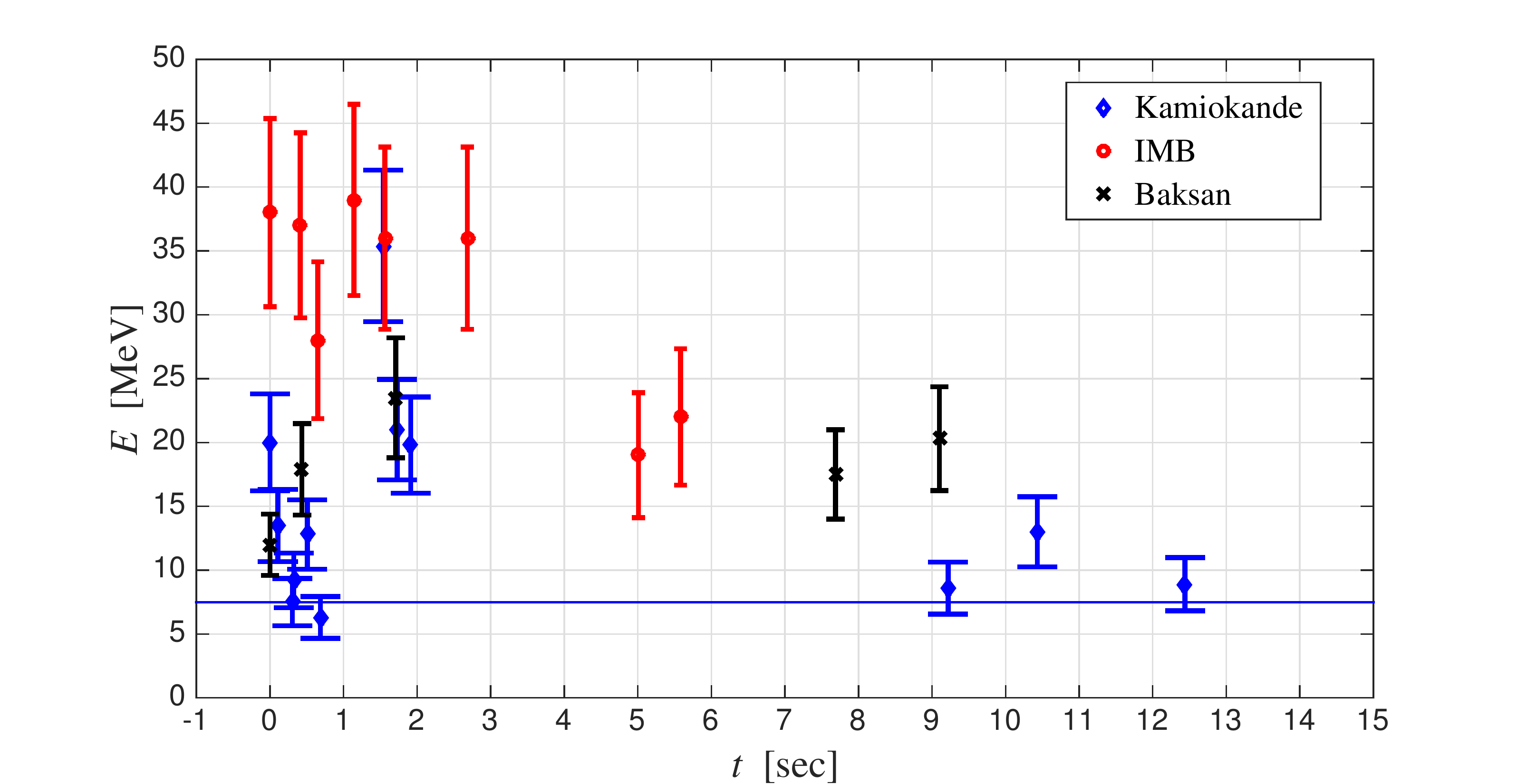}
\caption{Time series of the neutrino burst of SN1987A. Blue (diamond), Red (circle), and black (cross) markers denote reconstructed event energies for the Kamiokande, IMB, and Baksan detectors, respectively, with $1\sigma$ error bars. Horizontal blue line denotes the traditional 7.5~MeV threshold imposed in analyses of Kamiokande data. Note that the three time series are offset by unknown relative delays, likely of order 100~ms; here we set these delays to zero.
}
\label{fig:evts}
\end{figure}

Figure~\ref{fig:evts} depicts the time series of the SN1987A neutrino burst. Blue (diamond), Red (circle), and black (cross) markers denote reconstructed event energies for the Kamiokande~\citep{Hirata:1987hu}, IMB~\citep{Bionta:1987qt}, and Baksan~\citep{Alekseev:1987ej} detectors, respectively, with $1\sigma$ error bars. Horizontal blue line denotes the traditional 7.5~MeV threshold imposed in analyses of Kamiokande data. Note that the three time series are offset by unknown relative delays, likely of order 100~ms. Here we set these delays to zero; this has no impact on our results\footnote{We also note that Kamiokande observed four additional events at times 17.6,\,20.3,\,21.4, and 23.8~sec, with energies 6.5,\,5.4,\,4.6, and 6.5~MeV, respectively. These late-time events were below threshold for the original Kamiokande analysis. Nevertheless, they were included (together with proper background treatment) in the likelihood analysis of \citet{Loredo:2001rx}, \citet{Pagliaroli:2008ur}, and \citet{Ianni:2009bd}, and though we do not show them in~Figure~\ref{fig:evts} we include these events in our analysis too.}. 

Figure~\ref{fig:evts} shows signal events from IMB (that had, reportedly, no background) at $t>5$~sec. The Kamiokande event at $t=10.4$~sec, with $E_\nu>10$~MeV, also has only a small probability of being due to background~\citep{Loredo:2001rx}. For our purpose in this paper, the implication is that neutrino emission should last for at least 5-10~sec after core-collapse. Models for CITE that predict BH formation on time $t_{\textrm{BH}}\sim1-3$~sec, must invoke another neutrino source to replace the PNS cooling and accretion luminosity for $t>t_{\textrm{BH}}$. CITE can fulfil this requirement via accretion disc luminosity. Nevertheless, before going into more detail we can conclude that CITE models that predict RIAS formation on time $t_{\textrm{disc}}>5$~sec are excluded by the SN1987A neutrino data of IMB. It is thus important to investigate if reasonable stellar profiles can be found in which CITE operates successfully with an RIAS launch time $t_{\textrm{disc}}\approx5$~sec. We will tackle this task in Section~\ref{sec:numerical}.

Please look again at Figure~\ref{fig:evts}. Our main point in the rest of this section, is that the neutrino light curve is compatible with two different physical mechanisms accounting for the initial dense sequence of events on times $t\lesssim2-3$~sec, and the subsequent reduced luminosity on times $t>5$~sec. 
In fact, there is a time gap between the Kamiokande event at $t=1.9$~sec and the next Kamiokande events at $t>9$~sec. A comparably significant gap (the statement of significance requires modelling, that we provide later on) exists between the IMB event at $t=2.7$~sec and the next pair of events at $t=5$ and 5.6~sec. 

The time gaps in the neutrino data were noted in, e.g.,~\citet[][]{Spergel:1987ch}, \citet{Lattimer:1989}, and \citet{,Suzuki:1988qi}. \citet{Spergel:1987ch} commented on the possible hint for a discontinuity, but fitted a  continuous exponential PNS cooling model to the neutrino light-curve, finding a reasonable global fit. We have redone the analysis of~\citet{Spergel:1987ch} and we agree with their numbers. Indeed, the SN1987A neutrino data is too sparse for conclusive detailed modeling. However, it is important to note that~\citet{Spergel:1987ch}, and other analyses \citep[such as, e.g.,][]{Loredo:2001rx,Pagliaroli:2008ur,Ianni:2009bd}, did not have a theoretical model contender to the PNS accretion and cooling luminosity predicted within the neutrino mechanism. The situation for us is different. A time gap in the neutrino data, with intense PNS luminosity for $t\leq t_{\textrm{BH}}\sim1-3$~sec, silence for a few seconds, and renewed accretion disc luminosity, is precisely what we expect from CITE. 
In the next subsections we explore this point further with some statistical analyses. 

\subsection{A luminosity drop at $t\sim2$~sec?}\label{ssec:lumin1}
To obtain a basic assessment of the neutrino source luminosity, we make two simplifying assumptions:\\
(i) We assume that the neutrino distribution function at the source is a modified Fermi-Dirac spectrum with instantaneous temperature $T(t)$ and $\bar\nu_e$ luminosity $L_{\bar\nu_e}(t)$,
\be\label{eq:dNdE0}\frac{dN_{\bar\nu_e}^{(0)}}{dEdt}(t)=\frac{L_{\bar\nu_e}(t)}{c_L(\alpha)\,T^2(t)}\frac{\left(E/T(t)\right)^{2+\alpha}}{\exp\left(E/T(t)\right)+1},\ee
where $c_L(\alpha)=\left(1-2^{-3-\alpha}\right)\Gamma(4+\alpha)\zeta(4+\alpha)$. The mean $\bar\nu_e$ energy for this spectrum is\footnote{For reference, $c_L(0)\approx5.68$, $c_L(2)\approx118.26$, $c_T(0)\approx3.15$, $c_T(2)\approx5.07$.} $\langle E_\nu\rangle(t)=c_T(\alpha)\,T(t)$ with $c_T(\alpha)=c_L(\alpha)/c_L(\alpha-1)$. 
We set $\alpha=2$. The superscript on $dN^{(0)}_{\bar\nu_e}/dEdt$ denotes the spectrum at the source, before neutrino flavor mixing. \\
(ii) We neglect the contribution of $\bar\nu_\mu$ and $\bar\nu_\tau$ at the source. 
Using Eq.~(\ref{eq:dNdE0}) we compute the $\bar\nu_e$ differential flux at the detector,
\be\label{eq:Phi}\Phi_{\bar\nu_e}(t)=\frac{P_{ee}}{4\pi D_{SN}^2}\frac{dN_{\bar\nu_e}^{(0)}}{dEdt}(t),\ee
with the electron antineutrino survival probability $P_{ee}=0.67$ and with $D_{SN}=50$~kpc the distance to SN1987A. 

We perform a Poisson likelihood analysis for the Kamiokande, IMB, and Baksan neutrino data of SN1987A, including background and detector efficiency effects. We implement the analysis suggested in~\citet{Pagliaroli:2008ur} and in \citet{Ianni:2009bd} that modifies the method of~\cite{Loredo:2001rx} in the treatment of detector efficiency. We include only the dominant inverse-beta decay (IBD) reaction~\citep{Strumia:2003zx}. For detector efficiency and backgrounds, we use the updates given by~\cite{Vissani:2014doa}. 

Our first analysis of the data is as follows. We split the full neutrino event time series (16 events in Kamiokande, 8 events in IMB, and 5 events in Baksan) into eight time bins of equal log-space duration delimited by  $[0,0.25,0.5,1,2,4,8,16,32]$~sec. In each time bin, centred around the time $t_i$, we fix the source parameters $L_{\bar\nu_e}(t)$ and $T(t)$ to constant values $L_i,\,T_i$, independent from bin to bin. We then perform a bin by bin likelihood analysis in the two source parameters $L_i$ and $T_i$. 

\begin{figure}[htbp]
\includegraphics[width=1\linewidth]{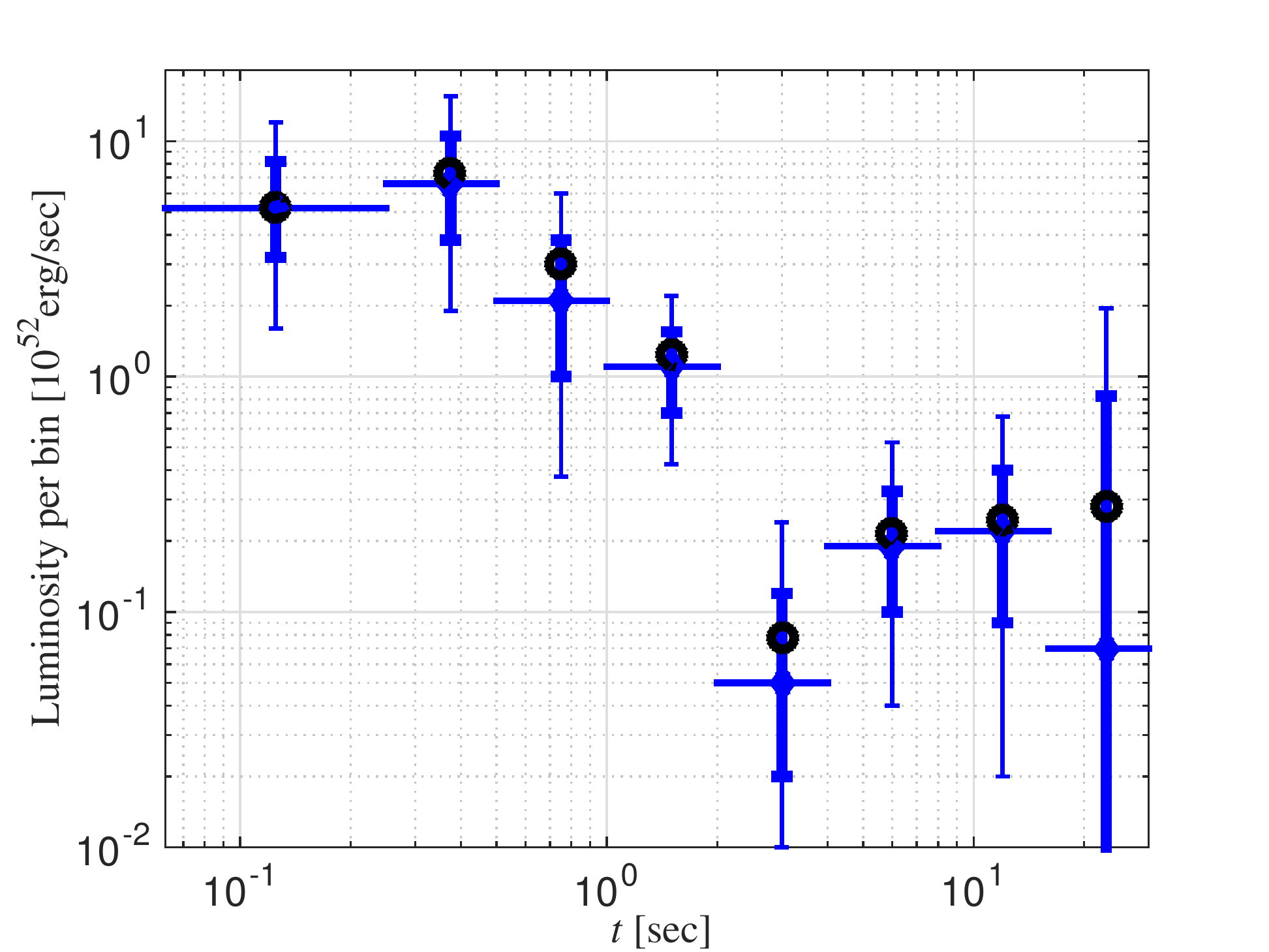}
\caption{Inferred $\bar\nu_e$ luminosity. Blue markers: Poisson likelihood analysis based on the simplified formulae Eqs.~(\ref{eq:dNdE0}-\ref{eq:Phi}), using the combined Kamiokande, IMB, and Baksan data, with thick (thin) error bars showing $1\sigma$ ($2\sigma$) ranges. Black circles: $R_\nu/E_\nu$ binned luminosity estimator from data; see Eq.~(\ref{eq:ideal}) and text around it. Note that (i) the luminosity estimator (black) does not account for detector background, while the Poisson fit (blue) subtracts it; (ii) omitting Baksan data does not affect the results significantly.}
\label{fig:lumin}
\end{figure}

Figure~\ref{fig:lumin} shows the result for the fitted source luminosity $L_i$. In each bin, the blue marker denotes the best fit luminosity, with thick (thin) vertical error bars obtained by fixing $T_i$ to its best fit point and letting $L_i$ vary within $\Delta\chi^2<1$ ($\Delta\chi^2<4$). (The black circles are explained in Appendix~\ref{app:EC}, and are not important for the discussion in this section.) The horizontal bars denote the time bin duration\footnote{We find it more informative for our current purpose to plot the instantaneous mean luminosity $L_{i}$ in each bin, rather than the energy per bin $t_iL_{i}$, despite the logarithmic bin assignment. We thank John Beacom for discussion on this point.}. The time gap in Figure~\ref{fig:evts} is reproduced in Figure~\ref{fig:lumin} as an order of magnitude drop in the $\bar\nu_e$ luminosity around $t\sim2$~sec. 

Regarding our parameterization in Eqs.~(\ref{eq:dNdE0}-\ref{eq:Phi}). Other assumptions about neutrino flavor mixing or source flavor composition can be made by reinterpreting the product $P_{ee}L_{\bar\nu_e}$. Our results are not affected significantly by the choice of $\alpha$. A plain thermal spectrum would have $\alpha=0$. Our choice of $\alpha=2$ applies if the dominant $\bar\nu_e$ source is $e^+n\to p\bar\nu_e$ from a plasma containing free nucleons and $e^\pm$ pairs, a reasonable scenario if the luminosity is dominated by accreting matter \citep[see, e.g.,][]{Perego:2015oqa}. Contributions due to $\bar\nu_\mu$ and $\bar\nu_\tau$ at the source are straightforward to include and do not affect our conclusions. Last, our parametrization of neutrino mixing ignores possibly important matter and neutrino self-induced oscillation effects \citep[see, e.g.,][]{Kartavtsev:2015eva,Mirizzi:2015eza}. Our choice of $P_{ee}=0.67\approx\cos^2\theta_{12}$ applies for normal mass hierarchy, with a strong matter effect causing adiabatic alignment of the $\bar\nu_e$ flavor state with the mass eigenstate $\bar\nu_1$~\citep{2002PhRvD..65g3008F}. We chose this treatment of flavor conversion mainly to facilitate comparison with previous work that used the same prescription~\citep{Loredo:2001rx,Pagliaroli:2008ur,Ianni:2009bd}.  

We now wish to compare different theoretical models for the supernova. In Section~\ref{ssec:lumin2} we follow~\citet{Loredo:2001rx} and \citet{Pagliaroli:2008ur} and perform a likelihood analysis of the neutrino mechanism using phenomenological models. In Section~\ref{ssec:lumin3} we devise analogues models, with the same number of free parameters, for the neutrino luminosity expected in CITE. We compare the performance of CITE models to that of the neutrino mechanism.

%%%%%%%%%%%%%%%%%%%%%%%
\subsection{Neutrino mechanism}
\label{ssec:lumin2}
Calculations within the neutrino mechanism suggest that the supernova explosion, if it is to occur at all, should occur within a few hundred milliseconds after core collapse \citep[see, e.g.,][]{OConnor:2011,Pejcha:2014wda}. Before the explosion, accretion of the stellar envelope onto the PNS produces accretion luminosity with nontrivial time dependence, that (for $\nu_e$ and $\bar\nu_e$) can dominate over the cooling luminosity of the PNS. However, after the explosion has cleared away the accreting matter, for post-bounce times $t>1$~sec, the neutrino mechanism has a robust prediction of continuous PNS cooling luminosity that is slowly decreasing with a characteristic time scale of a few seconds. 

\cite{Pagliaroli:2008ur} used phenomenological models of the neutrino flux to represent the predictions of the neutrino mechanism. The main models discussed there were: (i) simple exponential cooling of the PNS (EC model, details in Appendix~\ref{app:EC}), and (ii) exponential cooling + truncated accretion model (ECTA model, details in Appendix~\ref{app:EC}), with more free parameters added to describe the early accretion phase, that is truncated by hand at $t_{\textrm{accretion}}=0.55$~sec.

We have reproduced the best fit points for the ECTA and EC models, in agreement with~\cite{Pagliaroli:2008ur}. We also reproduce the Poisson likelihood difference $\Delta\chi^2\approx-10$ in favor of the ECTA model best-fit point as compared with that of the EC model. This statistical preference for the ECTA model led~\citet{Loredo:2001rx} and \citet{Pagliaroli:2008ur} to argue that the data provides evidence for an early accretion phase on time $t\lesssim0.5$~sec.

\citet{Loredo:2001rx} and \citet{Pagliaroli:2008ur} did not have a model that could address a sharp luminosity drop at time $t>1$~sec. It is interesting to re-inspect their results, keeping in mind CITE as an alternative theory. 
The neutrino time series as shown in Figure~\ref{fig:evts} has no events in either Kamiokande or IMB\footnote{There were no events in Baksan, either, for this time period.} during the time $t=2.7-5$~sec. We can use the best-fit EC and ECTA models of~\cite{Pagliaroli:2008ur} to calculate the Poisson probability for observing no events during this time. The results are given in Table~\ref{tab:1}. For the EC (ECTA) model, this probability is 0.4\% (2.2\%). 
\begin{table}[htp]
\caption{Poisson probability for the neutrino data during $t=2.7-5$~sec. For each theoretical model we calculate the number of signal events expected in each detector during this time. The Poisson probability for observing no events is given in the last column. IMB is assumed to have zero background. We take the threshold energy $E_{\textrm{th}}=4.5$~MeV for Kamiokande, with which we expect 0.43 background events during $t=2.7-5$~sec.}\label{tab:1}
\begin{center}
\begin{tabular}{|c|c|c|c|}
\hline
Model & Kamiokande & IMB & Poisson probability\\\hline\hline
EC ($\nu$ mechanism) & 3.65 & 1.56 & 0.4\%\\\hline
ECTA ($\nu$ mechanism) & 2.23 & 1.17 & 2.2\%\\\hline 
CITE & 0.03 & 0.01 & 62\%\\\hline
\end{tabular}
\end{center}
\label{default}
\end{table}

We learn that the time gap is somewhat unlikely from the point of view of the neutrino mechanism. In fact, we suspect that the improved global likelihood of the ECTA model, interpreted in~\citet{Loredo:2001rx} and in \citet{Pagliaroli:2008ur} as evidence for an early accretion phase, may actually be driven to some extent by the need to not over-shoot the late time luminosity gap at $t>2$~sec. To clarify this point, in Appendix~\ref{app:EC} we construct a binned Monte-Carlo (MC) analysis that provides time-dependent information on the performance of the fit.

\subsection{CITE: accretion for $\sim2-3$~sec, then black hole formation}
\label{ssec:lumin3}
The neutrino luminosity in CITE, for a progenitor star relevant to SN1987A and skipping the abrupt initial $\nu_e$ deleptonization burst that is unimportant for our purpose, follows three main phases.
\begin{enumerate}
\item PNS forms on core collapse, followed by accretion through a quasi-static accretion shock. This is the usual stalled shock of the core bounce. Electron-flavor neutrino luminosity is dominated by nucleon conversion reactions with $L_{\nu_e}\approx L_{\bar\nu_e}$, where $L_{\bar\nu_e}\sim\frac{GM_{PNS}\dot M_{PNS}}{2R_{PNS}}\sim10^{52}\left(\frac{M_{PNS}}{2~M_{\odot}}\right)\left(\frac{\dot M_{PNS}}{0.1~M_{\odot}/{\rm sec}}\right)\left(\frac{25~{\rm km}}{R_{PNS}}\right)$~erg/sec. PNS cooling produces additional luminosity $L_x\sim(0.3-0.6)L_{\bar\nu_e}$, with $L_{\nu_\mu}\approx L_{\bar\nu_\mu}\approx L_{\nu_\tau}\approx L_{\bar\nu_\tau}\equiv L_x$. State-of-the art examples are presented in~\citet{Perego:2015oqa} and in \citet{Mirizzi:2015eza}. The accretion phase lasts for 1-3~sec after bounce, until the PNS accumulates baryonic mass $\mysim2-3~M_{\odot}$ (depending on details of the EOS~\citep{OConnor:2011}) leading to BH formation

\item BH forms, absorbing the matter downstream to the accretion shock on a time scale of miliseconds. Spherical accretion directly onto the BH produces small neutrino luminosity, $L_{\bar \nu_e}\lesssim10^{47}$~erg/sec, because the accreting matter in the absence of a shock does not have time to radiate its gravitational binding energy before it goes through the horizon (or, from the perspective of an observer at infinity, gets redshifted to nothing). Thus, BH formation leads to an abrupt cut-off in the neutrino luminosity

\item A quiescent phase, corresponding to quasi-spherical accretion on the BH, should begin at $t_{\textrm{BH}}$ and last for $\mysim1-10$ seconds. However, for CITE to work~\citep{Kushnir:2015mca}, angular momentum in the envelope must produce a centrifugal barrier, leading to an accretion disc and to the launch of the RIAS that propagates outward and eventually triggers the explosion. Matter in the disc at the base of the RIAS heats up and re-initiates neutrino emission, dominated again by nucleon conversion reactions and potentially reaching $\mysim10^{51}$~erg/sec.  
\end{enumerate}

We construct a toy phenomenological parametrization for the neutrino luminosity of SN1987A in CITE, in the spirit of the neutrino mechanism analysis of~\citet{Loredo:2001rx} and \citet{Pagliaroli:2008ur}. We use six free  parameters, the same number of parameters as used in~\citet{Loredo:2001rx} and in \citet{Pagliaroli:2008ur} to define the ECTA model of the neutrino mechanism.  

Two of our parameters define the basic CITE time scales: $t_{\textrm{BH}}$ denoting BH formation, and  $t_{\textrm{disc}}$ denoting accretion disc formation (and launch of the RIAS). 
To model the PNS accretion phase (phase 1 above), we build on the simulations of~\citet{Perego:2015oqa} for their HC19.2 pre-supernova progenitor model, when their artificial trigger \citep[denoted "PUSH" in][]{Perego:2015oqa} is not used to start an explosion. We define two parameters, $f_L$ and $f_E$. During the interval $0<t<0.8$~sec, where~\cite{Perego:2015oqa} provided numerical results, our model luminosity is $L_{\bar\nu_e}(t)=f_L\times L_{\bar\nu_e}^{HC19.2}(t)$, $L_x(t)=f_L\times L_x^{HC19.2}(t)$, with mean neutrino energy $\langle E_{\bar\nu_e}\rangle(t)=f_E\times \langle E_{\bar\nu_e}\rangle^{HC19.2}(t)$, $\langle E_x\rangle(t)=f_E\times \langle E_x\rangle^{HC19.2}(t)$. Here, $L_{\bar\nu_e}^{HC19.2}$ is the $\bar\nu_e$ luminosity reported by~\cite{Perego:2015oqa}, etc. For $0.8~{\rm sec}<t<t_{\textrm{BH}}$ \citep[where][did not provide numerical results]{Perego:2015oqa}, we let the luminosity decrease as $L_{\bar\nu_e}(t),\,L_x(t)\propto1/t$, while the energies are set to rise linearly $\langle E_{\bar\nu_e}\rangle(t),\,\langle E_x\rangle(t)\propto t$. For $t>t_{\textrm{BH}}$, we set $L_x(t)=0$, and $L_{\bar\nu_e}(t)=\frac{2L_{disc}}{1+t/t_{\textrm{disc}}}\left(1-e^{-\left(t/t_{\textrm{disc}}\right)^{k}}\right)$ with $k=100$. This form gives a fast rise for the accretion disc $\bar\nu_e$ luminosity, consistent with what we find in our numerical simulations in the next section. The mean energy during the disc phase is $\langle E_{\bar\nu_e}\rangle(t)=\frac{2E_{disc}}{1+t/t_{\textrm{disc}}}$. 

To summarize, our six free parameters are (1) $t_{\textrm{BH}}$ and (2) $t_{\textrm{disc}}$, denoting BH and subsequent disc formation times; (3) $f_L$ and (4) $f_E$, constant factors by which we modulate the numerical results for the HC19.2 SN1987A progenitor model of~\cite{Perego:2015oqa}, to obtain the luminosity before BH formation; and (5) $L_{disc}$ and (6) $E_{disc}$, characterizing the late neutrino emission of the accretion disc around the BH. 

Calculating the likelihood for our CITE parametrization, we find several configurations with Poisson likelihood superior to the best fit ECTA model of~\cite{Pagliaroli:2008ur}. For instance, the following model
\be\label{eq:modelcite}
&{\rm CITE:}\\
&t_{\textrm{BH}}=2.7~{\rm sec},\,t_{\textrm{disc}}=5~{\rm sec},\,f_L=0.67,\,f_E=0.56,\no\\
&L_{disc}=4\times10^{51}~{\rm erg/sec},\,E_{disc}=15~{\rm MeV},\no\ee
has $\Delta\chi^2$ smaller by 6.8 compared to the ECTA model of the neutrino mechanism. For comparison, within the neutrino mechanism, the ECTA model has $\Delta\chi^2$ smaller by 9.8 than that of the EC model, the latter having 3 parameters less; this was considered in~\cite{Loredo:2001rx} and in \citet{Pagliaroli:2008ur} as evidence for an accretion phase.

Our CITE model is obviously consistent with no events during the time $t=2.7-5$~sec. We give the Poisson probability in Table~\ref{tab:1}. Comparing to the neutrino mechanism, the $2.7-5$~sec time gap is the source for the improved likelihood of CITE. In Appendix~\ref{app:EC} we repeat our binned MC analysis for the model in~(\ref{eq:modelcite}). Incidentally, as we have based the first second of our CITE model light-curve on the  non-exploding numerical simulation of~\cite{Perego:2015oqa}, it is safe to say that the early time neutrino data does not require a transition between accretion luminosity to PNS cooling luminosity. Continued accretion is consistent with the data.  

Finally we comment on the fit results in~(\ref{eq:modelcite}). 
\begin{itemize}
\item The value of $t_{\textrm{BH}}=2.7$~sec is in good agreement with progenitor models as in~\cite{Perego:2015oqa}. We believe that $t_{\textrm{BH}}\sim1-3$~sec is a robust prediction of CITE for strong explosions, as can be seen form analytical estimates as well as numerical simulations~\citep{OConnor:2011}.
\item We view the requirement $t_{\textrm{disc}}=5$~sec as an observational constraint on CITE, at least when attempting to interpret SN1987A IMB data. We analyze the implications of this constraint further in the next section. We do not see anything particularly un-natural with $t_{\textrm{disc}}$ of a few seconds, as long as CITE can operate successfully. However, we should stress that while the formation of the disc, by itself, is a built-in ingredient in CITE, the precise timing $t_{\textrm{disc}}=5$~sec we deduce here from the neutrino data is not a generic prediction of the model: as seen in \cite{Kushnir:2015mca}, CITE could operate just as well with $t_{\textrm{disc}}>10$~sec. This is to be contrasted with the more robust prediction of $t_{\textrm{BH}}$ in the previous item.
\item The values of $f_L$ and $f_E$ we find correspond to moderate modulation of the results of~\cite{Perego:2015oqa}. Much larger modulations could arise from varying the input pre-collapse profile within observational constraints. 
\item Last, the best fit disc neutrino energy $E_{disc}$ is higher by about a factor of 2, and the best fit disc neutrino luminosity $L_{disc}$ is higher by about an order of a magnitude, than our estimate of the disc emission in the next section. However, these parameters are not tightly constrained by the data. For example, keeping the other parameters at the same value as in~(\ref{eq:modelcite}), but reducing $L_{disc}$ to $10^{51}~{\rm erg/sec}$, gives Poisson likelihood for CITE that is worse by $\Delta\chi^2\approx6.2$ compared to the $L_{disc}=4\times10^{51}~{\rm erg/sec}$ of~(\ref{eq:modelcite}), but still improved by $\Delta\chi^2\approx0.6$ compared to the analogues ECTA model of the neutrino mechanism. Varying both $L_{disc}$ and $E_{disc}$ within $\Delta\chi^2=4$ around the reference values in~(\ref{eq:modelcite}) we find values in the range $L_{disc}\sim(1-10)\times10^{51}~{\rm erg/sec}$ and $E_{disc}\sim10-20$~MeV, with higher $L_{disc}$ correlated with lower $E_{disc}$ and vice-verse.
\end{itemize}

%%%%%%%%%%%%%%%%%%%
\section{Numerical simulations}
\label{sec:numerical}

In this section we perform 2D numerical simulations of CITE. We have two goals:\\ 
(i) to verify that CITE can operate with RIAS launch time $t_{\textrm{disc}}\approx5$~sec, reproducing $E_{\textrm{kin}}$ and $M_{\textrm{Ni}}$ in the ballpark of observations for SN1987A.\\ 
(ii) to study the accretion disc neutrino luminosity relevant for CITE on times $t>t_{\textrm{disc}}$.

With respect to item (ii), we stress that our calculations are preliminary. Our code is Newtonian and our treatment of neutrino transport is simplistic. More sophisticated codes exist in the literature \citep[see, e.g.,][]{Perego:2015oqa,Mirizzi:2015eza,OConnor:2011}; our estimates here motivate the application of these tools to the scenario of CITE. Beyond the technical limitations of the simulation, the problem of the neutrino luminosity of BH accretion discs suffers from theoretical uncertainties due to the implementation of viscosity. Here we set the viscosity to zero. Estimates for different assumptions of the viscosity can be found in~\citet{Popham:1998ab} and in \citet{MacFadyen:1998vz}. 
Because of these limitations, we do not attempt to reproduce the neutrino light curve in any detail besides from the rough luminosity and time scales.

We aim to simulate the process of the accretion disc at times $t\gtrsim3$~sec and the subsequent CITE, and we do not attempt to reproduce the early phase of PNS and BH formation \citep[again see, e.g.,][for details of this early phase]{Perego:2015oqa,Mirizzi:2015eza,OConnor:2011}. We assume that once sufficient mass, $M>2-3~M_{\odot}$, has accreted through the inner boundary of our simulation, $r_{\rm inner}$ (to be specified below), the central object forms a BH. Before this time, the flow below $r\sim10^7$~cm in our simulation does not capture correctly the standing shock above the PNS; however, for $t>t_{\textrm{BH}}$ the shocked material is quickly absorbed in the BH and by $t>4$~sec -- still many dynamical times prior to disc formation in our simulation -- we expect that our calculation provides a reasonable approximation of the flow down to $r$ near the last stable orbit.

\subsection{Pre-collapse profile}\label{ssec:prof}

We use the same methods from \cite{Kushnir:2015mca}, so here we only highlight a few aspects of the simulations. We do not simulate the collapse at $r<r_{\rm inner}$ and details of the progenitor on $r<r_{\rm inner}$ are unimportant for the results. On $r>r_{\rm inner}$ the pre-collapse profile is composed of shells with constant entropy per unit mass, $s$, constant composition, and in hydrostatic equilibrium. We place $1.6\,M_{\odot}$ within $r<2\cdot10^{8}\,\textrm{cm}$, representing a degenerate iron core. This choice roughly reproduces the PNS mass in~\cite{Perego:2015oqa} for their HC19.2 progenitor model when PUSH is not used to trigger an explosion. The region between $r_{\textrm{inner}}$ and $r=2\cdot10^{8}\,\textrm{cm}$ is filled with $s=1\,k_{B}$  iron (in hydrostatic equilibrium), which is prevented from burning. This prescription is chosen for simplicity, and we defer more detailed analysis to future work. Note that the region inwards of $r=2\cdot10^8$~cm falls through $r_{\rm inner}$ in time $t\approx\pi\sqrt{r^3/2GM(r)}\approx0.4$~sec after core collapse, so the composition in this region has a negligible effect on the results of the simulation.

\begin{figure}[htbp]
\includegraphics[width=1\linewidth]{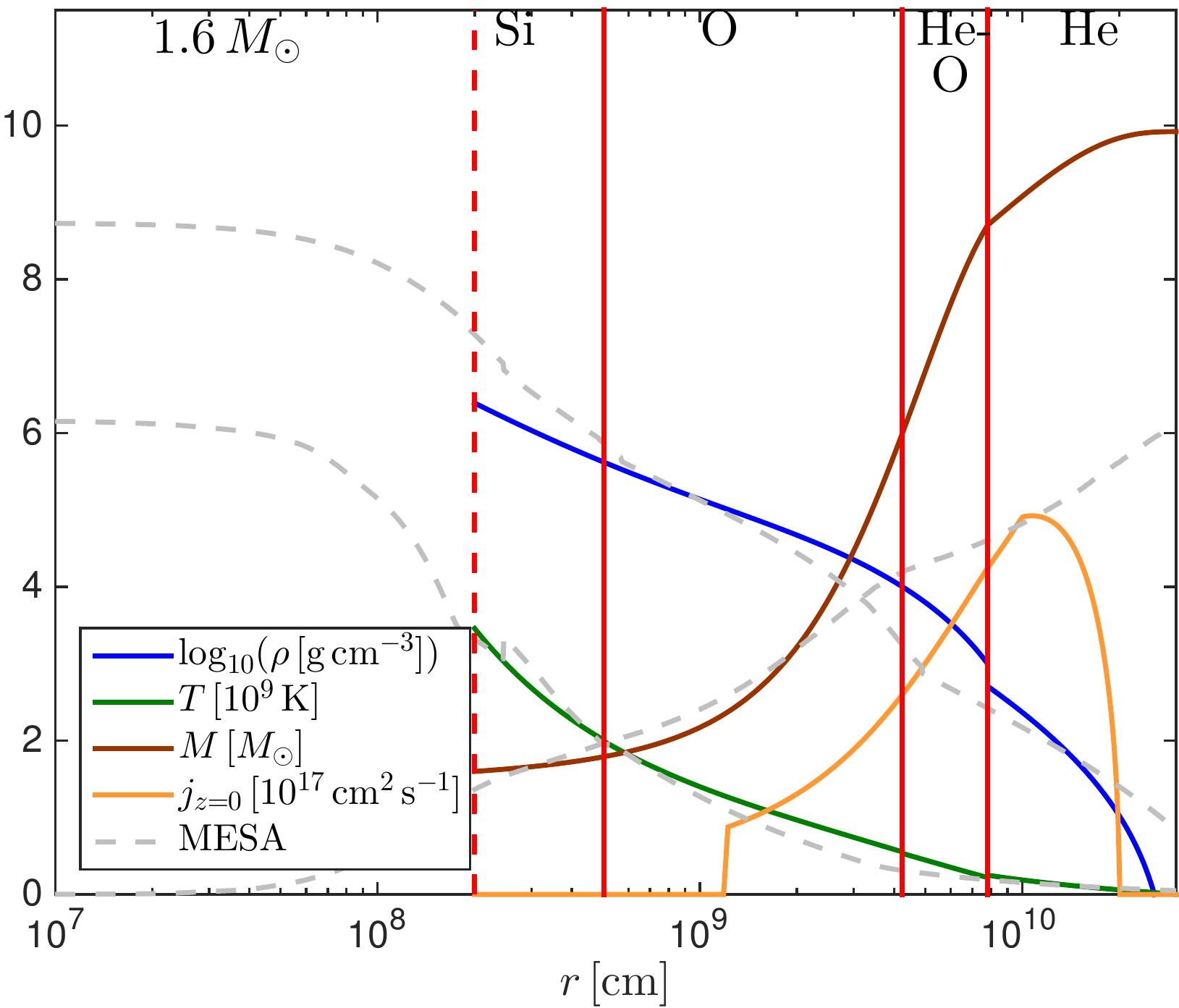}
\caption{Pre-collapse stellar profile  (density, temperature, enclosed mass, and specific angular momentum on the equatorial plane, $j_{z=0}$) used in CITE simulation (the profile below $r=2\cdot10^{8}\,\textrm{cm}$ has negligible effect on the results, see text for details). The density, temperature, and enclosed mass profiles are similar to pre-collapse profiles of a $20\,M_{\odot}$ star (dashed gray), calculated by Roni Waldman with MESA \citep[][]{MESA}.}
\label{fig:profile}
\end{figure}

The base of the He-O shell is placed at a mass coordinate of $6\,M_{\odot}$, a radius of $4.25\cdot10^{9}\,\textrm{cm}$, and a density of $10^{4}\,\textrm{g}\,\textrm{cm}^{-3}$. The shell is composed of equal mass fraction of helium and oxygen. The local burning time at the base of the shell is $\myapprox\,700\,\textrm{s}$, which is $100$ times the free-fall time at this position. The total mass of the He-O shell is $\myapprox\,2.7\,M_{\odot}$. Pure oxygen (helium) is placed below (above) the He-O shell.  Oxygen is replaced with silicon where $T>2\cdot 10^9\,\textrm{K}$, to prevent fast initial burning. The angular momentum is initially distributed such that $\frot$, the ratio of the centrifugal force to the component of the gravitational force perpendicular to the rotation axis, is constant $\frotz=0.02$ throughout the profile, except for the following:
\begin{itemize}
\item $\frot=0$ at $r<1.2\cdot10^{9}\,\textrm{cm}$.
\item $\frot=0$ at large radii, and increases linearly with decreasing radius between $r=2\cdot10^{10}\,\textrm{cm}$ and $r=10^{10}\,\textrm{cm}$ to $\frotz$. This is done for numerical stability, and has a small effect on the results.
\end{itemize}
The stellar profile used in the analysis is shown in Figure~\ref{fig:profile}.

The stellar profile defined here is designed to achieve $t_{\textrm{disc}}\approx5$~sec and disc neutrino luminosity $L_{\bar\nu_e}\sim10^{51}$~erg/sec. 
An estimate of the disc formation time is given by
\be\label{eq:an1} t_{\textrm{disc}}\approx2t_{\textrm{ff}}(r_f)=\pi\sqrt{\frac{r_f^3}{2GM(r_f)}}.\ee
Here $t_{\textrm{ff}}(r)$ is the free-fall time at pre-collapse radial coordinate $r$, $M(r)$ is the enclosed mass, and $r_f$ is the radial coordinate on the $z=0$ plane where the centrifugal force fraction $f$ first becomes greater than zero. The factor of 2 in Eq.~(\ref{eq:an1}) sums (i) the (almost) free-fall trajectory of the mass element initially at $r_f$ down to the disc formation radius $r_{disc}\approx(f/2)r_f\ll r_f$, and (ii) the time it takes the rarefaction wave starting at core-collapse to propagate out from $r=0$ to $r_f$ \cite[Because the initial profile is in hydrostatic equilibrium, this sound travel time is again roughly equal to the free-fall time at $r_f$;][]{Kushnir:2014oca}. In Figure~\ref{fig:profile}, $r_f=1.2\cdot10^9$~cm and $M(r_f)=2.34~M_{\odot}$, so we estimate $t_{\textrm{disc}}\approx5.2$~sec.

The disc neutrino luminosity can be estimated by the gravitational binding energy accreting through the disc,
\be\label{eq:an2} L_{\bar\nu_e}&\sim&\frac{GM_{disc}\dot M_{disc}}{2r_{disc}}\\
&\approx& 10^{51}~\left[\frac{(f/2)\,r_f}{10^7~{\rm cm}}\right]^{-1}\left[\frac{M(r_f)}{2M_{\odot}}\right]\left[\frac{\dot M_{disc}}{0.05M_{\odot}/{\rm sec}}\right]~{\rm erg/sec}.\no\ee
This estimate assumes that half the disc emission is in $\bar\nu_e$. We scaled the  mass accretion rate through the disc by a typical value. 

Note that Eqs.~(\ref{eq:an1}-\ref{eq:an2}) are only used to tune the initial stellar profile before running the numerical simulations. We do not use these estimates in the numerical calculations described next.

\subsection{Simulations and results}\label{ssec:res}
The problem considered is axisymmetric, allowing the use of two-dimensional numerical simulations with high resolution. We employ the FLASH4.0 code with thermonuclear burning \citep[Eulerian, adaptive mesh refinement][]{Fryxell:2000} using cylindrical coordinates $(R,z)$ to calculate one quadrant, with angular momentum implementation as in \citet[][]{Kushnir:2015mca}. Layers below the inner boundary, $r_{\textrm{inner}}$, are assumed to have already collapsed, and the pressure within this radius is held at zero throughout the simulation. We assume that neutrinos escape freely through the outer layers. 

\begin{figure}[htbp]
\includegraphics[width=\linewidth]{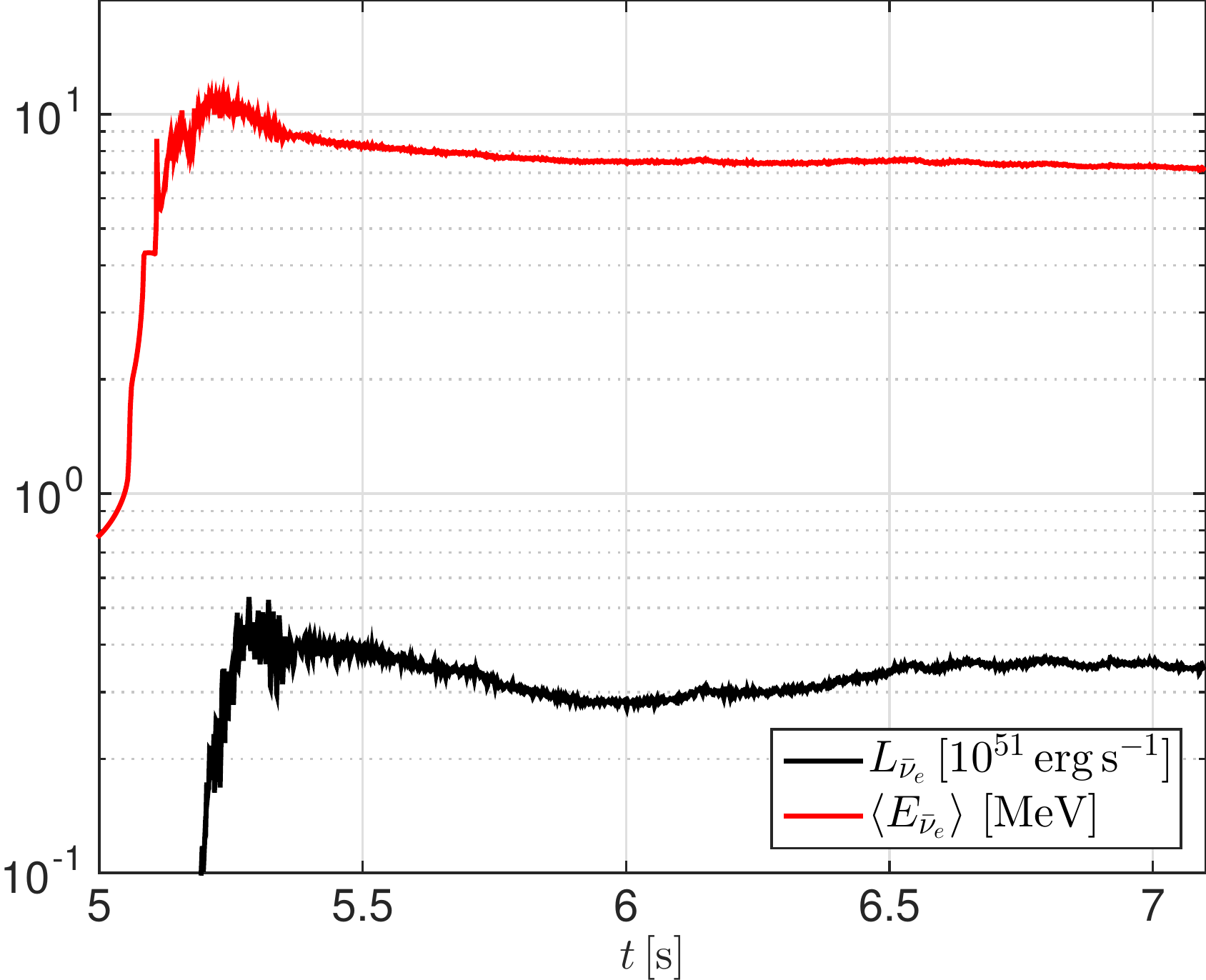}
\caption{Mean energy (red) and luminosity (black) for $\bar\nu_e$ taken from the numerical simulation.}
\label{fig:Num1}
\end{figure}

We perform two different simulation runs based on the same stellar profile. 
\begin{enumerate}
\item First, the thermonuclear explosion was calculated with $r_{\textrm{inner}}=60\,\textrm{km}$, a resolution (i.e. minimal allowed cell size within the most resolved regions) of $\myapprox14\,\textrm{km}$ and a 13-isotope $\alpha$-chain reaction network (similar to the APPROX13 network supplied with FLASH with slightly updated rates for specific reactions, especially fixing a typo for the reaction $^{28}\textrm{Si}(\alpha,\gamma)^{32}\textrm{S}$, which reduced the reaction rate by a factor $\myapprox4$). This setup is sufficient for calculating the disc formation, RIAS launch, and the resulting thermonuclear explosion. An ignition of a detonation wave was obtained at $t\approx25\,\textrm{sec}$, which resulted in an explosion with $E_{\textrm{kin}}\approx6\cdot10^{50}\,\textrm{erg}$ and $M_{\textrm{Ni}}\approx0.035\,M_{\odot}$. Both of these values are in the ballpark of, though smaller by a factor of $\myapprox2-3$ than the observed values of SN1987A~\citep{Utrobin:2011}. 
\item Second, to calculate the neutrino light curve we used $r_{\textrm{inner}}=30\,\textrm{km}$, a resolution of $\myapprox2\,\textrm{km}$ and the APPROX19 reaction network (to allow helium disintegration to nucleons). The required high resolution and small value of $r_{\textrm{inner}}$ allowed us to continue the calculation for only a few seconds after the disc formed. The nucleon conversion rates were estimated by $\myapprox9\cdot10^{23}\left(T/10^{11}\,\textrm{K}\right)^{6}X_{n}\,\textrm{erg}\,\textrm{s}^{-1}\,\textrm{g}^{-1}$  \citep{Qian:1996}, where $X_{n}$ is the mass fraction of neutrons. The baryonic mass below $r_{\textrm{inner}}$ reached $2\,M_{\odot}$ at $\myapprox2.5\,\textrm{sec}$ and the RIAS formed at $t_{\textrm{disc}}\approx5\,\textrm{sec}$, increasing $L_{\bar{\nu}_{e}}$ to $\approx5\cdot10^{50}\,\textrm{erg}\,\textrm{s}^{-1}$ where the mean energy of the neutrinos\footnote{Mean neutrino energy was approximated from the matter temperature, averaged by neutrino emissivity and assuming $\alpha=2$ in Eq.~(\ref{eq:dNdE0}).} is estimated by $\left\langle E_{\bar{\nu}_{e}}\right\rangle\approx10\,\textrm{MeV}$ (see Figure~\ref{fig:Num1}).  
\end{enumerate}

\begin{figure}[htbp]
\includegraphics[width=\linewidth]{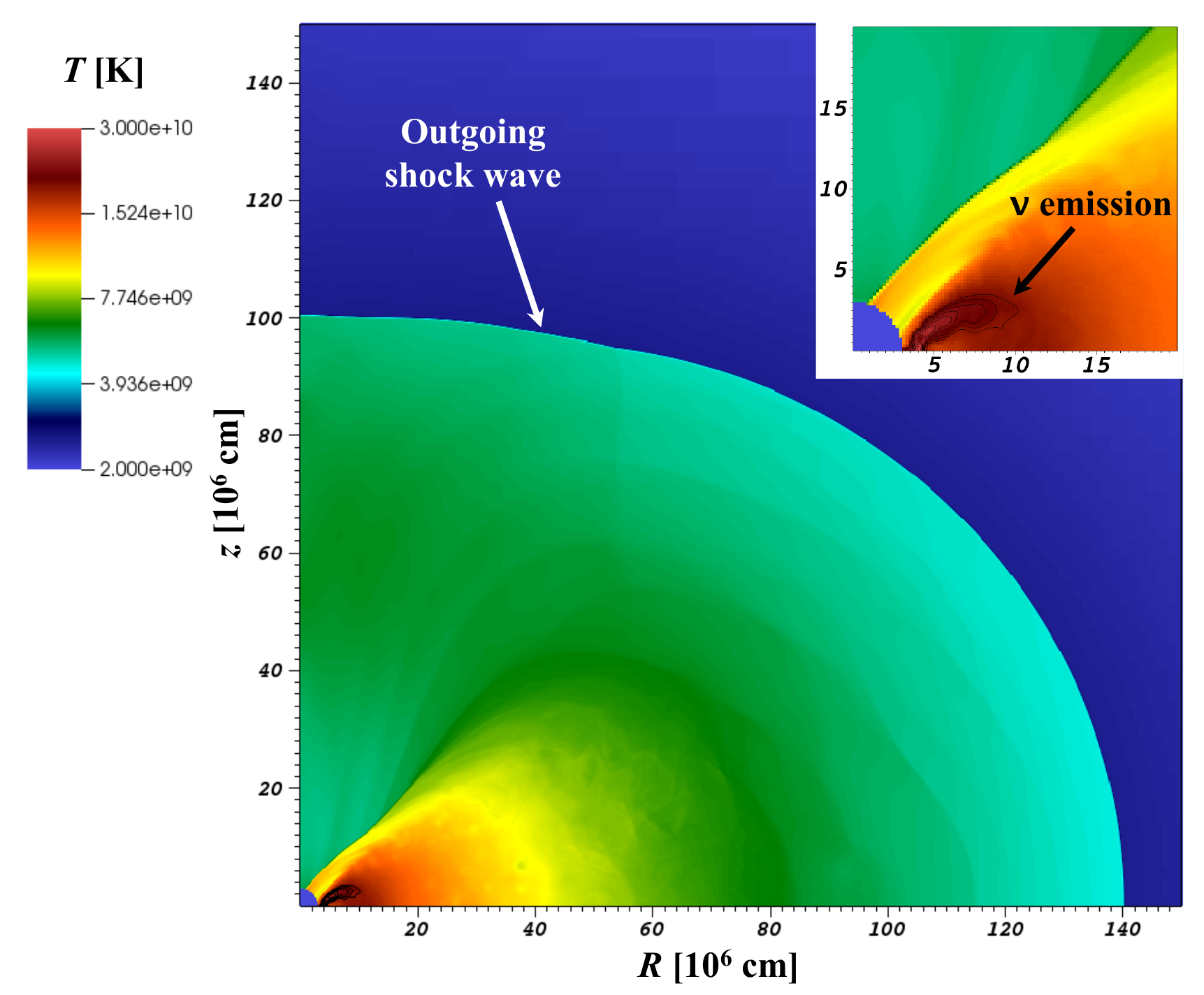}
\caption{BH accretion disc neutrino luminosity in CITE. Logarithmic temperature map at $5.5\,\textrm{s}$ since collapse with neutrino emission contours (black, 10 contours logarithmically distributed between $3\cdot10^{19}-3\cdot10^{20}\,\textrm{erg}\,\textrm{s}^{-1}\,\textrm{g}^{-1}$). The inset shows a zoomed map around the neutrino emission region.}
\label{fig:Num2}
\end{figure}

A snapshot of the disc and RIAS at time $5.5\,\textrm{sec}$ is shown in Figure~\ref{fig:Num2}. The neutrino emission originates from radii $30-100\,\textrm{km}$, but mostly dominated from $30-40\,\textrm{km}$ where the typical densities are $\textrm{few}\times10^{9}\,\textrm{g}\,\textrm{cm}^{-3}$.

Increasing the resolution to $\myapprox1\,\textrm{km}$ changes the results by less than $10\%$, but increasing $r_{\textrm{inner}}$ to $40\,\textrm{km}$ leads to a reduced luminosity by $30-40\%$ and reduced energies by $10\%$. We conservatively estimate that our results are accurate to only a factor of a few, as our simulations are Newtonian and velocities of $\mysim0.5c$ are achieved near the emission region. Furthermore, the Schwartzchild radius of the central BH at this time is $R_s\approx10$~km, so our disc, that ignores general relativistic effects, is located not far above the last stable circular orbit. Nevertheless, our results demonstrate that  $L_{\bar{\nu}_{e}}\sim10^{51}\,\textrm{erg}\,\textrm{s}^{-1}$ with $E_{\bar\nu_e}\sim10$~MeV is possible for $t>t_{\textrm{disc}}$. 

Our results can be compared to those of~\citet{Popham:1998ab} and of \citet{MacFadyen:1998vz} in the context of the collapsar model, that shared a similar setup to ours. The latter included a free parameter to account for viscosity effects, finding accretion disc neutrino luminosity with a range encompassing our result here.  

%%%%%%%%%%%%%%%%%%%
\section{Conclusions}
\label{sec:conclusion}

The neutrino burst of SN1987A has been traditionally used to advocate for the neutrino mechanism operating in exploding CCSNe. Our goal in this paper was to give a first analysis of the neutrino signal expected in CITE, as a competing mechanism of CCSNe, and to compare it to the SN1987A signal. 
The questions we addressed and our results were as follows.

There is a common claim in the literature that direct BH formation would be incompatible with SN1987A. This claim is usually based on two arguments: (i) the neutrino mechanism predicts a NS remnant, and/or (ii) direct BH formation would cut-off the neutrino emission, leaving the signal events at $t>5$~sec unexplained.

We find that this claim is, at least currently, unjustified. First, the neutrino mechanism has not yet been shown to operate successfully and reproduce the observations of SN1987A. Therefore its failure is not a good cause to exclude BH formation. CITE provides one potential counter example. Second, if the progenitor of SN1987A possessed a rotating envelope then an accretion disc would form around the BH. Such accretion discs are known to be copious neutrino emitters and could explain the late-time neutrino events of SN1987A.

In Section~\ref{sec:data}  we gave a statistical analysis of the neutrino emission in CITE, along the lines used by~\citet{Loredo:2001rx} to study the neutrino mechanism. While the statistical significance of such analysis is limited by the sparse data, we find that: (i) there is a hint in the data for a luminosity drop around $t\sim2$~sec, right in the ballpark where CITE predicts BH formation; (ii) the neutrino mechanism is in some tension with this luminosity drop, while CITE could address it naturally.

The neutrino events at $t>5$~sec imply that CITE should be operative with RIAS formation as early as that. This is a nontrivial constraint that was not considered in~\citet{Kushnir:2015mca}. It can be summarized by Eqs.~(\ref{eq:an1}-\ref{eq:an2}) with $t_{\textrm{disc}}\approx5$~sec and $L_{\bar\nu_e}\approx10^{51}$~erg/sec.  
In Section~\ref{sec:numerical} we performed 2D numerical simulations guided by these constraints. Without yet attempting a systematic survey of possible profiles, we were able to find such profile that yields an explosion in the rough ballpark of the observations ($E_{\textrm{kin}}$ about a factor of 3 and $M_{\textrm{Ni}}$ about a factor of 2 below that of SN1987A). Further study of different initial profiles is needed to derive more conclusive results. We also gave order of magnitude estimates of the luminosity and typical energy of the neutrino emission produced by the BH accretion disc at the base of the RIAS, finding results on the low side, but not inconsistent with the data.

We close with comments on further work.
\begin{itemize}
\item We are eager to see independent simulations of CITE, to compare with the work of \citet{Kushnir:2015mca}. In particular, it is important to investigate whether the pre-collapse initial conditions required for CITE can be obtained with stellar evolution models.
\item Many particle physics analyses used the neutrino burst of SN1987A to constrain new physics beyond the Standard Model, such as axions or sterile neutrinos (see, e.g.,~\citet{1996slfp.book.....R}). Most of these works assumed PNS cooling luminosity, as suggested within the neutrino mechanism. Our results here imply that these analyses may need to be revisited. 
\item If CITE works in nature, then the neutrino burst of the next strong Galactic CCSN may give us front-row seats to the formation of an event horizon in real time with current neutrino detectors. This may have already happened, albeit with limited statistics, with SN1987A. Access to phenomena near the event-horizon motivates construction of a few Megaton neutrino detector that will observe extragalactic CCSNe on a yearly basis \citep{2011PhRvD..83l3008K}.
\end{itemize}

\acknowledgments
We thank John Beacom, Avishay Gal-Yam, Juna Kollmeier, and Fukugita Masataka for discussions, Kohta Murase for early collaboration, and Boaz Katz, Eli Waxman and Yosef Nir for comments on the manuscript. D.~K. gratefully acknowledges support from the Friends of the Institute for Advanced Study. FLASH was in part developed by the DOE NNSA-ASC OASCR Flash Center at the University of Chicago.

\appendix
\section{Modelling of the neutrino mechanism, and Monte Carlo study}\label{app:EC}
We recap here some details on the EC and ECTA models of~\citet{Loredo:2001rx} and \citet{Pagliaroli:2008ur}. We also describe a Monte Carlo (MC) analysis designed to clarify the time dependence in the different models. 

The EC model has 3 free parameters for the neutrino source: (i) NS initial temperature $T_c$, (ii) NS cooling time $\tau_c$, and (iii) NS neutrinosphere radius $R_c$. The $\bar\nu$ luminosity is assumed to scale as $L_{\bar\nu_e}(t)\propto R_c^2\,T^4_c(t)$ with $T_c(t)=T_ce^{-\frac{t}{4\tau_c}}$. In addition to the source parameters, three unknown time shifts between the zero of time in the three detectors Kamiokande, IMB, and Baksan are also marginalized over, with the best-fit model of~\cite{Pagliaroli:2008ur} corresponding to these time shifts being zero. 
In addition to direct $\bar\nu_e$ emission, equal luminosity is assumed to be emitted in $\mu$ and $\tau$ flavors, $L_{\bar\nu_e}=L_{\bar\nu_\mu}=L_{\bar\nu_\tau}\equiv L_x$. The $x$-flavor temperature is set by hand to 1.2 times the $\bar\nu_e$ temperature. To account for $L_x$ in terms of the effective $L_{\bar\nu_e}$ of Eq.~(\ref{eq:Phi}), we convert $L_{\bar\nu_e}^{\rm effective}=L_{\bar\nu_e}^{\rm model}+(1/P_{ee}-1)L_{\bar\nu_x}^{\rm model}$, with $L_{\bar\nu_x}=L_{\bar\nu_\mu}=L_{\bar\nu_\tau}$.

The ECTA model adds, on top of the 3 parameters of EC, 3 more free parameters intended to describe an early accretion phase preceding the explosion: (iv) accretion temperature $T_a$, (v) accretion time scale $\tau_a$, and (vi) a parameter $\mu$ proportional to the over-all accretion luminosity. \cite{Pagliaroli:2008ur} assumes that the accretion luminosity consists purely of $e$ flavor, setting $L_x=0$ during the accretion phase and turning it back on once accretion is stopped and replaced by NS cooling as above. In addition to the free parameters (iv-vi), time dependence for the accretion luminosity, $L_x\sim L_{x,0}/(1+t/0.5~{\rm sec})$, is introduced in \citet{Loredo:2001rx} and in \citet{Pagliaroli:2008ur} without counting the functional form or the time scale of 0.5~sec as another free parameter (instead, the 0.5~sec time scale is argued to arise in numerical simulations).

We move on to describe our binned MC procedure. In the limit that energy-dependent detector efficiency and background are not important, a good proxy for the source luminosity during some time interval $\Delta t$ is given by the sum of event inverse-energy,
\be \frac{R_\nu}{E_\nu}\equiv \frac{1}{\Delta t}\sum_k\frac{1}{E_k},\ee
where the sum goes over the neutrino events detected during $\Delta t$. To see this, note that the detection cross section at the relevant neutrino energies ($8~{\rm MeV}<E_\nu<45~{\rm MeV}$) can be approximated by $\sigma(E_\nu)\approx\bar\sigma\left(E_\nu/{\rm MeV}\right)^2$, with $\bar\sigma=6.8\times10^{-44}$~cm$^2$. Ignoring background and energy-dependent detector efficiency, we can compute the expected value of $R_\nu/E_\nu$ given source luminosity $L_{\bar \nu_e}$ (constant in time during $\Delta t$),
\be\label{eq:ideal}\left\langle\frac{R_\nu}{E_\nu}\right\rangle&\approx&\frac{N_p\,P_{ee}}{4\pi D_{SN}^2}\int dE\frac{\sigma(E)}{E}\,\frac{dN_{\bar\nu_e}^{(0)}}{dEdt}=\left(\frac{N_p\,P_{ee}\,\bar\sigma}{4\pi D_{SN}^2}\right)L_{\bar\nu_e},\ee
where $N_p$ is the effective number of target protons in the detector. For an ideal detector we have $L_{\bar\nu_e}\approx\left(\frac{10^{32}}{N_p}\right)\left(\frac{\left\langle R_\nu/E_\nu\right\rangle}{\rm MeV^{-1}sec^{-1}}\right)\times10^{53}~{\rm erg/sec}$. In practice energy-dependent efficiency introduces an effective low-energy threshold that lowers the proportionality coefficient on the RHS of Eq.~(\ref{eq:ideal}) in a detector-dependent way. In addition, a small correction is introduced due to the small mismatch between the incoming neutrino energy and the reconstructed positron energy in the IBD detection event. Combining all three detectors, we find that the replacement $N_p\to 0.29\left(N_{p,\rm Kam}+N_{p,\rm IMB}+N_{p,\rm Bak}\right)=1.8\times10^{32}$ in Eq.~(\ref{eq:ideal}) for the luminosity estimator
\be
\label{eq:estimator}
L_{\bar\nu_e}^{R_{\nu}/E_{\nu}}\approx\left(\frac{R_\nu/E_\nu}{\rm MeV^{-1}sec^{-1}}\right)\times5.6\cdot10^{52}~{\rm erg/sec},
\ee
reproduces the data well for the source parameter $\alpha=2$ adopted in most of this section\footnote{Changing to $\alpha=0$ we find for the scaling factor $0.29\to0.25$. None of our results is affected significantly.}. 
The result of applying Eq.~(\ref{eq:estimator}) to the data is shown by black markers in Figure~\ref{fig:lumin}. Note that the luminosity estimator in Eq.~(\ref{eq:estimator}) does not account for detector background, while the Poisson fit (blue in Figure~\ref{fig:lumin}) automatically subtracts it. 

\begin{figure}[htbp]
\includegraphics[width=0.45\linewidth]{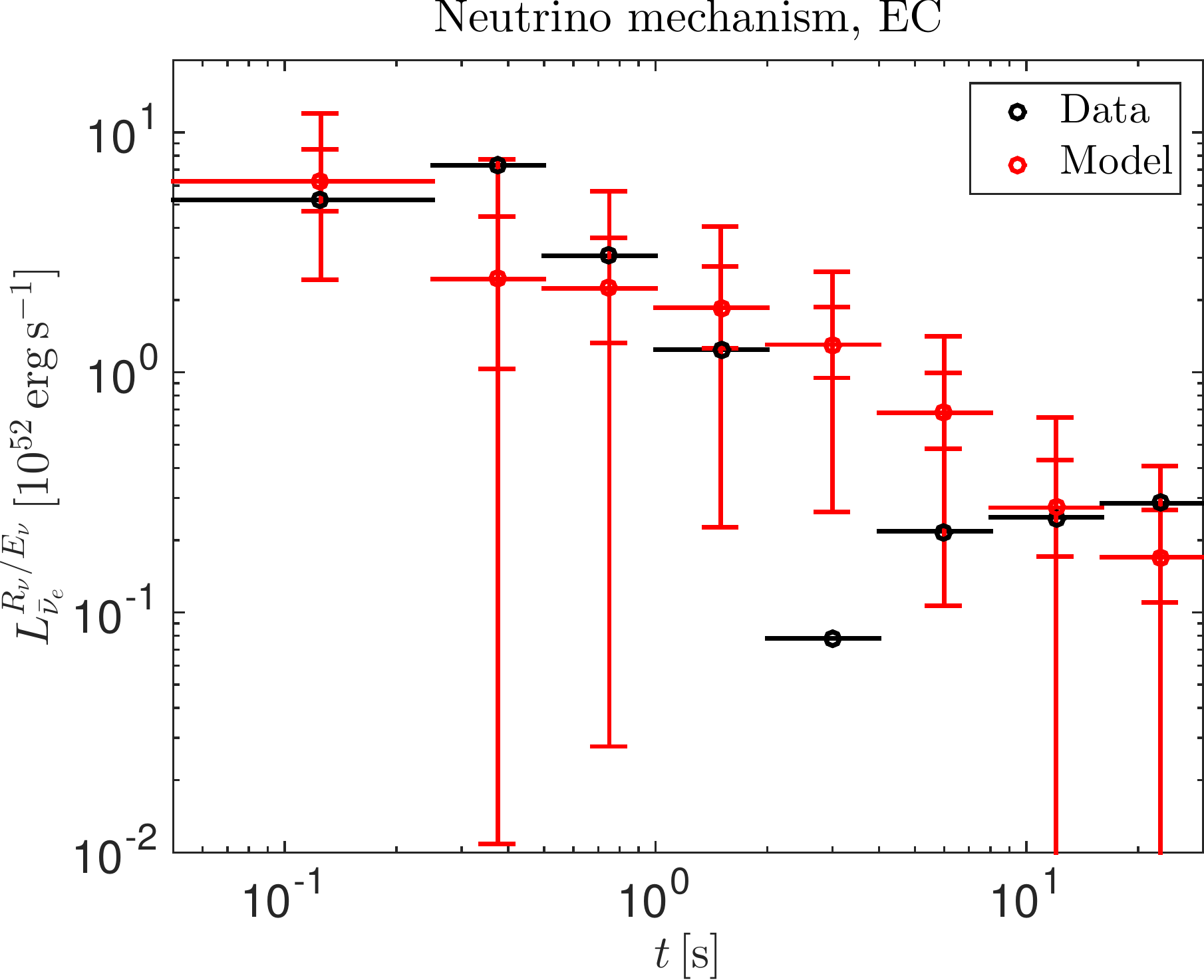}
\includegraphics[width=0.45\linewidth]{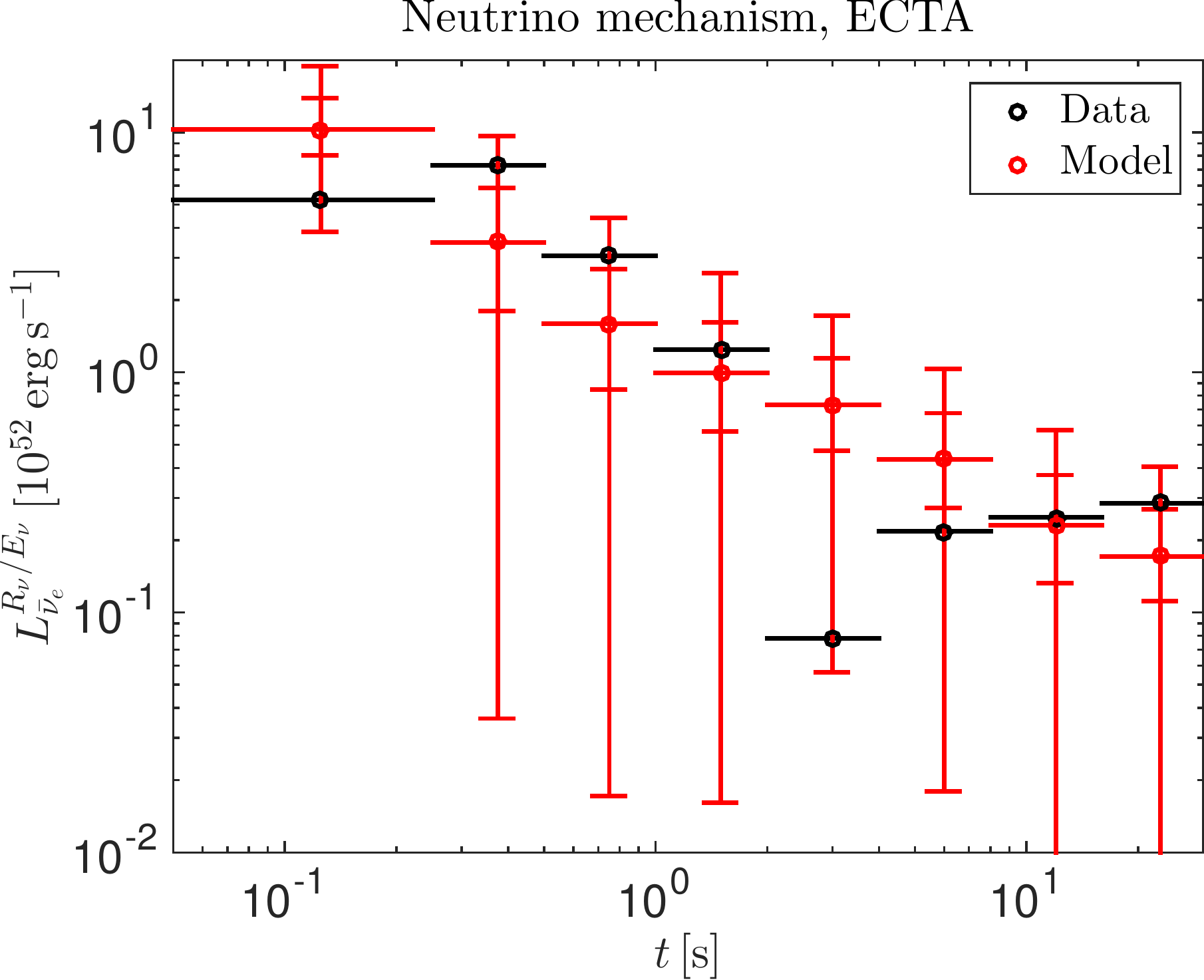}
\caption{MC distribution of the binned $L_{\bar\nu_e}^{R_{\nu}/E_{\nu}}$ source luminosity estimator. Left: EC model. Right: ECTA model. }
\label{fig:R2E_EC_ECTA}
\end{figure}

Armed with our quick-to-compute luminosity estimator $L_{\bar\nu_e}^{R_{\nu}/E_{\nu}}$ from Eq.~(\ref{eq:estimator}), we  generate mock data samples and compute the distribution of $L_{\bar\nu_e}^{R_{\nu}/E_{\nu}}$ in different time bins. In Figure~\ref{fig:R2E_EC_ECTA} we show the Monte Carlo (MC) results for $L_{\bar\nu_e}^{R_{\nu}/E_{\nu}}$ (red markers), computed for the best fit EC (left) and ECTA (right) models of~\cite{Pagliaroli:2008ur}. The MC results we show are converged to a few percent with $5\cdot10^{4}$ mock samples. 

Figure~\ref{fig:R2E_EC_ECTA} suggests that much of the statistical tension associated with the simple EC PNS cooling model is driven by the luminosity drop at $t\sim2$~sec. In order to not  overshoot the event rate during this time, the EC model is forced to low luminosity on earlier times, leading to tension in the $t\sim0.25-0.5$~sec time bin. The ECTA model can fix some of this tension, raising the luminosity at $t\lesssim0.5$~sec while using extra free parameters to keep the late time ``cooling PNS" luminosity not too high.  

\begin{figure}[htbp]
\includegraphics[width=0.45\linewidth]{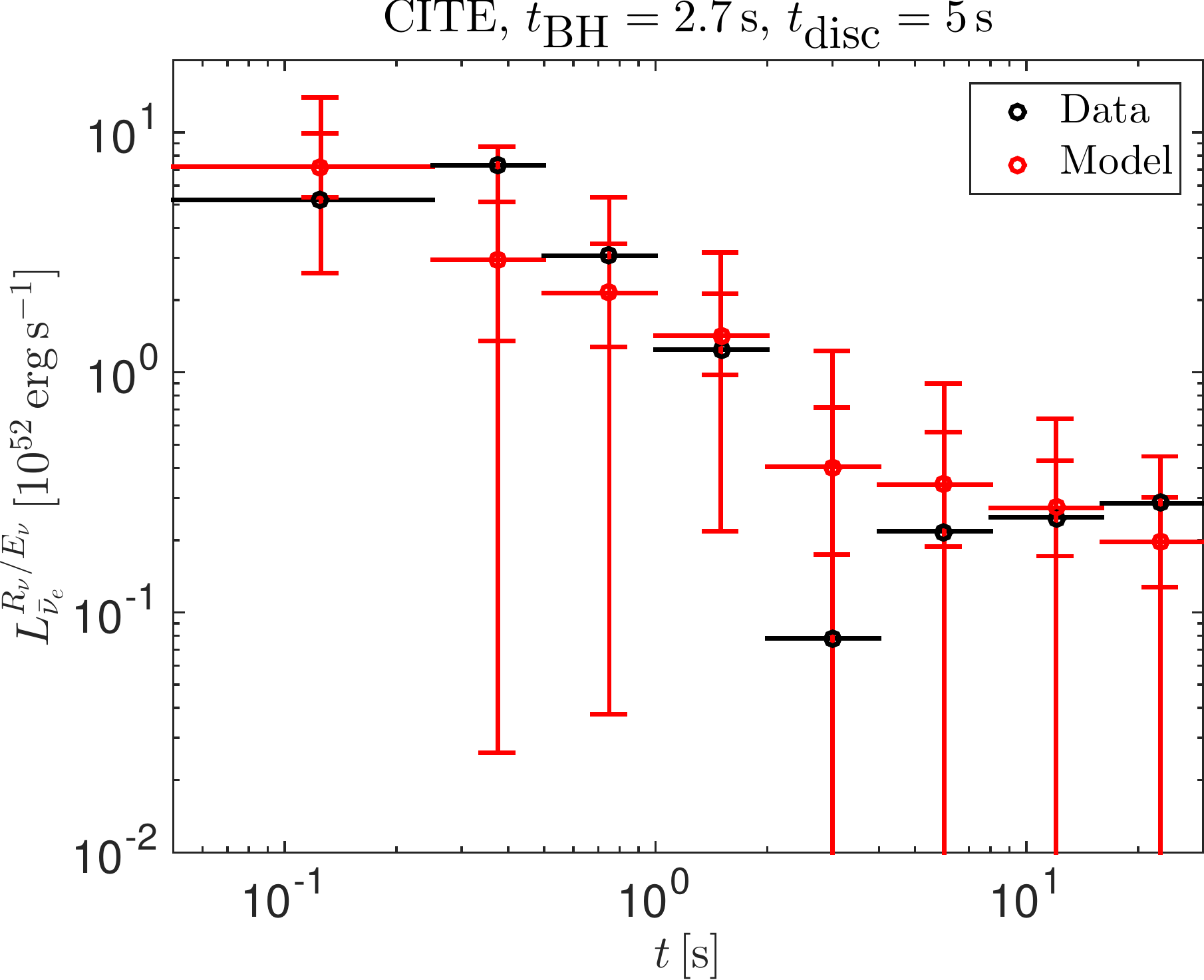}
\caption{MC distribution of the binned $L_{\bar\nu_e}^{R_{\nu}/E_{\nu}}$ source luminosity estimator for CITE, as parameterized in Eq.~(\ref{eq:modelcite}).}
\label{fig:R2E_CITE}
\end{figure}

In Figure~\ref{fig:R2E_CITE} we repeat our MC procedure for the $L_{\bar\nu_e}^{R_{\nu}/E_{\nu}}$ binned luminosity estimator in CITE, using the model of Eq.~(\ref{eq:modelcite}). Compared with the ECTA model, we find somewhat improved consistency with the data.

\end{document}